\documentclass[prb,twocolumn,showpacs,amsmath,amssymb,superscriptaddress]{revtex4}
\usepackage[english]{babel}
\usepackage{graphicx}
\newcommand{\ket}[1]{|{#1}\rangle}
\newcommand{\ave}[1]{\langle{#1}\rangle}
\newcommand{\innx}[3]{\langle{\vphantom{#3}#1}|{#2}|{\vphantom{#1}#3}\rangle}
\newcommand{\innxx}[3]{\left\langle{\vphantom{#3}#1}\right|{#2}\left|{\vphantom{#1}#3}\right\rangle}
\begin{document}
\title{Spin-polaron excitations in a doped Shastry-Sutherland model}
\author{S. \surname{El Shawish}}
\affiliation{J. Stefan Institute, SI-1000 Ljubljana, Slovenia}
\author{J. \surname{Bon\v ca}}
\affiliation{Faculty of Mathematics and Physics, University of
Ljubljana, SI-1000 Ljubljana, Slovenia} \affiliation{J. Stefan
Institute, SI-1000 Ljubljana, Slovenia}
\date{\today}
\begin{abstract}
Using variational algorithm on an infinite Shastry-Sutherland (SS)
lattice we show that the introduction of a static nonmagnetic impurity
into a dimerized ground state leads to a formation of a small,
localized spin polaron surrounding the impurity site. Due to a
particular symmetry of the SS lattice, the polaron is extremely
anisotropic with a short spatial extent. The presence of nonmagnetic
impurities leads to a formation of pronounced in-gap peaks in the
dynamical spin structure factor, which we attribute to the
spin-doublet excitations of a single unpaired spin $S=1/2$ surrounded
by triplet fluctuations. Our results are relevant for the description
of SrCu$_2$(BO$_{3}$)$_{2}$ compound when doped with nonmagnetic atoms
at Cu sites.
\end{abstract}
\maketitle

\section{Introduction}
The relation between disordered spin liquids with a gap in the
spin excitation spectrum and superconductivity has aroused a lot
of interest. It was suggested \cite{i-I-sc} that doping a gapped
system may lead to hole pairing and superconductivity. In this
sense the idea of finding a superconducting state in a doped
two-dimensional spin-liquid SrCu$_{2}$(BO$_{3}$)$_{2}$ compound,
representing the  first physical realization of the
Shastry-Sutherland (SS) model,
\cite{p2-smith91,i-topleq2,p2-miyahara03} has been proposed
\cite{s-kumar,s-kimura,p4-chung} due to its structural similarity
with the high-temperature cuprates.
So far, numerical
calculations do not seem to support superconductivity in the doped
SS model. \cite{p4-leung}

Another aspect of the study of the doped SrCu$_{2}$(BO$_{3}$)$_{2}$ is
the influence of impurities on the nature of the singlet (dimerized)
ground state. When $J'/J$ of the underlying SS model is close to but
below the critical value $\sim0.7$, \cite{p2-miyahara03} at finite
doping, the system may select other ordered or disordered states,
which may in turn strongly affect the magnetism of this material. It
is rather straightforward to realize that replacing magnetic Cu ions
with nonmagnetic ones breaks the singlet nature of the bonds, thus
providing an additional frustration, resulting in the reduction of the
spin gap. The effects of doping should materialize in inelastic
neutron scattering through an enhanced lifetimes of single-triplet
excitations -- magnons. A finite lifetime associated with magnon
scattering on spin excitations surrounding the nonmagnetic ions would
in turn lead to a broadening of the single-triplet peak line in the
doped SrCu$_{2}$(BO$_{3}$)$_{2}$.

The SS model consists of a relatively simple spin ($S=1/2$)
Hamiltonian,
\begin{equation}
H=J\sum_{\ave{i,j}}{\bf S}_i\cdot {\bf S}_j+
J'\sum_{\ave{i,j}'}{\bf S}_i\cdot {\bf S}_j; \label{eq_shs0}
\end{equation}
here $\ave{i,j}$ denotes the nearest neighbors (NN) on bonds $J$
(dimers) and $\ave{i,j}'$ next-nearest neighbors (NNN) on bonds
$J'$. We introduce a single nonmagnetic impurity -- a site with a
missing spin, into the model in Eq.~(\ref{eq_shs0}). It is well
known that for $J'/J\lesssim 0.7$ the exact ground state can be
written as a product of singlet dimers. \cite{p2-miyahara03} We
should stress that the introduction of a single hole into the
exact dimerized ground state has nontrivial consequences.

Previous theoretical studies have been focused mostly on the
doping of SS model with mobile holes,
\cite{s-kumar,p4-vojta,p4-leung,p4-chung} with the intention to
explore the hole-pairing mechanism relevant for superconductivity.
In our work we limit our investigations to one static nonmagnetic
impurity (magnetic polaron) within the SS lattice. Our
calculations are relevant to the case where the substitution with
nonmagnetic ions takes place at Cu sites. We use a variational
method within a variational space constructed by successively
applying the off-diagonal parts of Hamiltonian~(\ref{eq_shs0}) on
the starting (0th--generation) state vector. The method yields
converged results, valid within the infinite system, for static
quantities such as the value of the spin gap as well as for the
spin distribution function. We also explore the properties of the
first few low-lying excited spin-polaron states. Finally, we
compute the dynamical spin structure factor as a function of the
wavevector.

For simplicity, we have decided to neglect the anisotropic
Dzyaloshinsky-Moriya interactions originating in spin-orbit coupling
\cite{dmint} and used the (bare) SS Hamiltonian (\ref{eq_shs0}).


%
\section{Variational Approach in 1-Hole Limit}
%

\begin{figure}[!t]\begin{center}
    \begin{minipage}[b]{0.22\textwidth}\begin{center}
    \includegraphics[width=\textwidth]{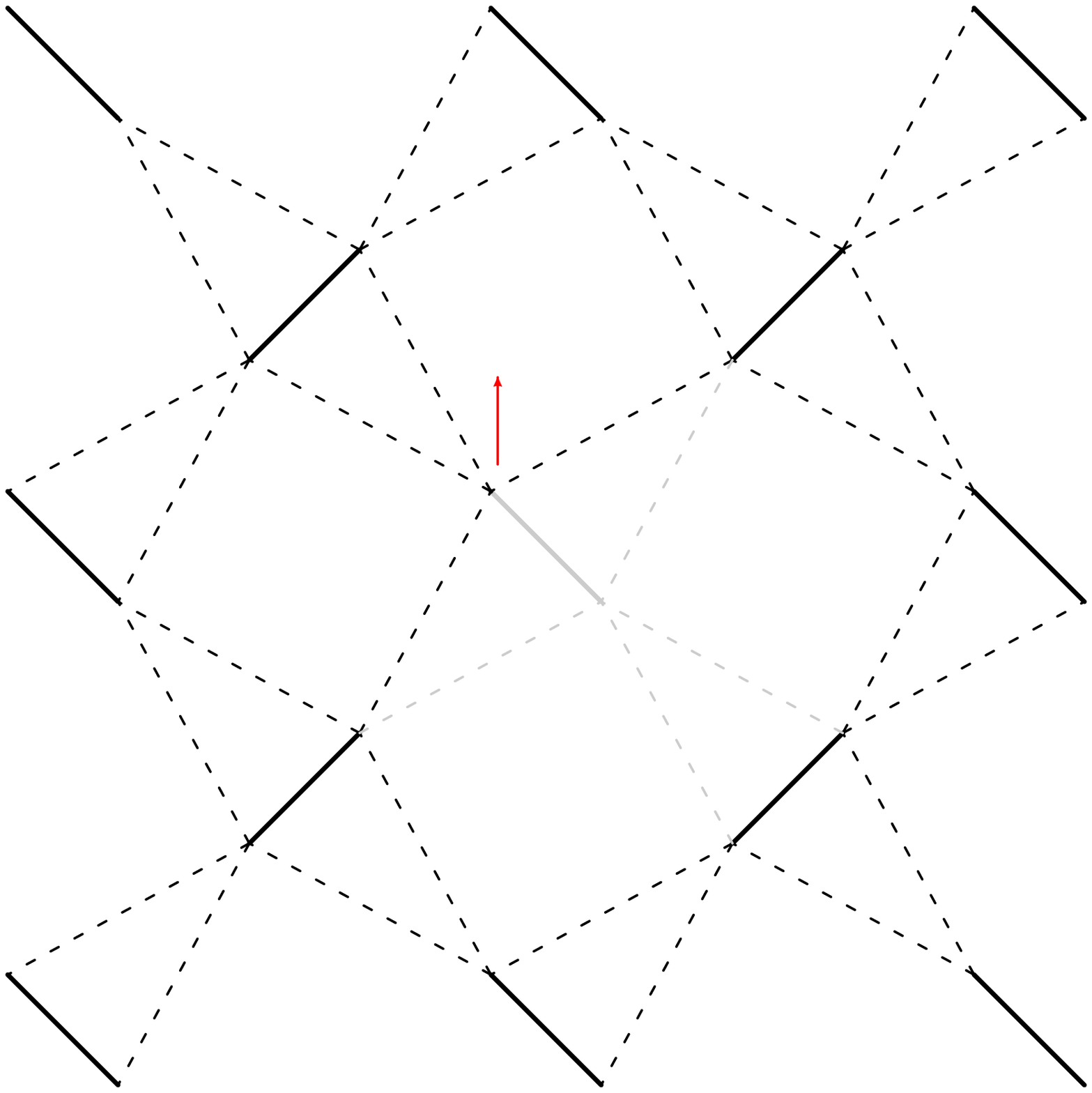} (a)
    \end{center}\end{minipage} \hskip0.5cm
    \begin{minipage}[b]{0.22\textwidth}\begin{center}
    \includegraphics[width=\textwidth]{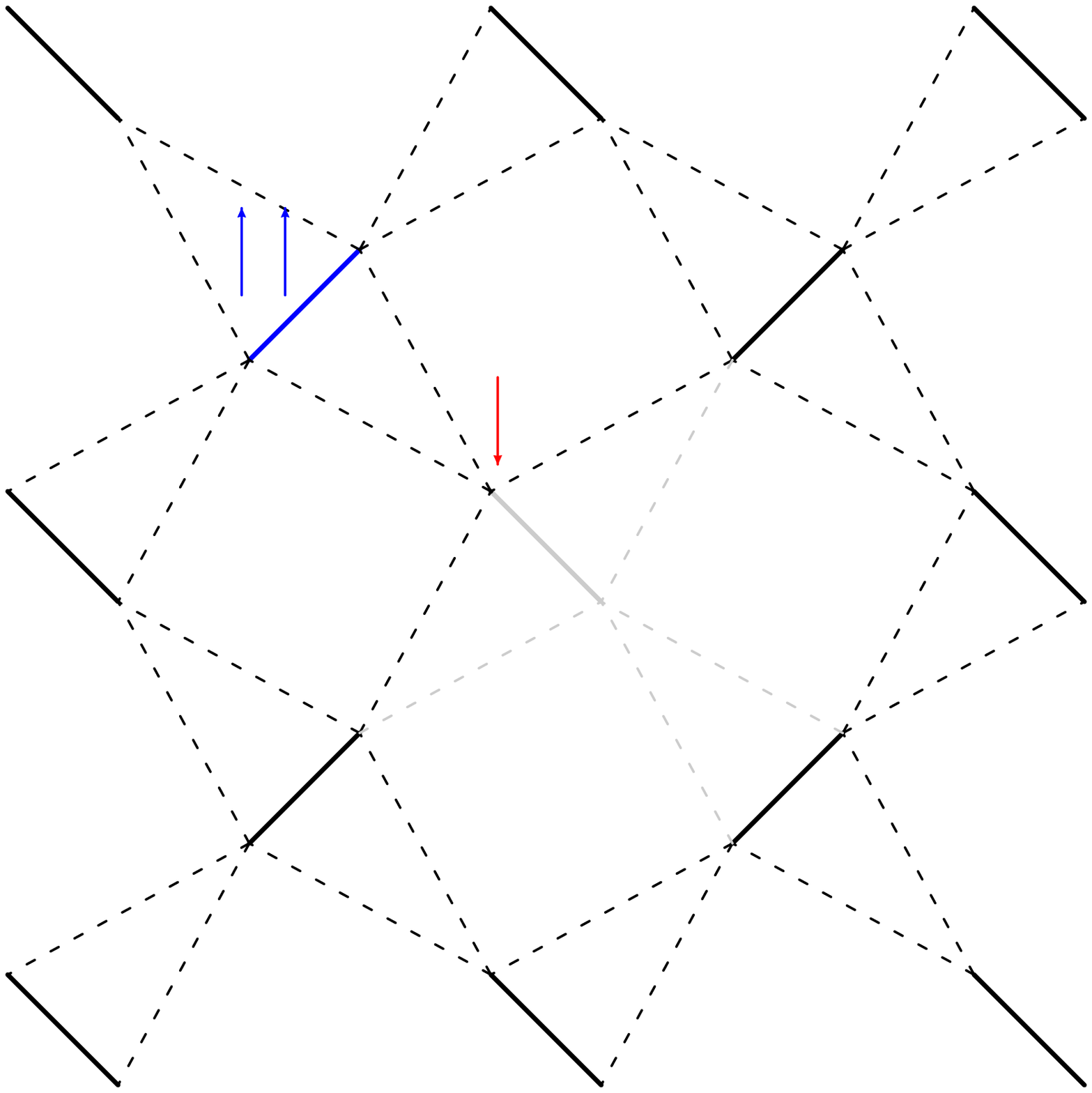} (b)
    \end{center}\end{minipage}
    \caption{(color online) Two initial states ($M=0$) in the 1-hole
    variational procedure defined on infinite SS lattice: (a) state
    $\ket{1_0}$ with uncoupled spin $S_{\rm u}^z=1/2$ ($\uparrow$)
    within a sea of singlet dimers (black solid lines), and (b) state
    $\ket{2_0}$ with $S_{\rm u}^z=-1/2$ ($\downarrow$) and one $S^z=1$
    triplet dimer ($\uparrow\uparrow$). Dashed black lines denote
    off-diagonal terms ($J'$) of $H$ (\ref{eq_shs0}).}
\label{fig_doped}
\end{center}\end{figure}

A variational space was constructed by starting from the
0th-generation state vector $\ket{1_0}$ representing one doped hole
with a neighboring uncoupled spin within a sea of singlet dimers
(Fig.~\ref{fig_doped}(a)). First generation, $M=1$, of variational
states was consequently obtained by acting with $H$ (\ref{eq_shs0}) on
state $\ket{1_0}$. Successive generations leading to $M$th order
states $\{\ket{i_M}\}$ were obtained starting from all previously
generated ones $\{\ket{i_{M-1}}\}$. A state can be uniquely determined
by the orientation of the spin adjacent the impurity site, $S_{\rm
u}^z=\pm1/2$, and the arrangement of $S^z=0$, $+1$ and $-1$ triplet
dimers in the vicinity of the hole, $\{{\bf r}_i^0\}$, $\{{\bf
r}_i^{+1}\}$ and $\{{\bf r}_i^{-1}\}$, respectively. With ${\bf
r}_i^{S^z}$ we denote the positions of triplet-dimer centers relative
to the impurity site.
{$i$-th variational state within $M$th generation}
is thus written as
\begin{equation}
   \ket{i_M}\equiv\ket{S_{\rm u}^z,\{{\bf r}_i^0\},\{{\bf r}_i^{+1}\},
   \{{\bf r}_i^{-1}\}} \label{eq_p4-var}
\end{equation}
where a fixed ordering of ${\bf r}_i^{S^z}$ vectors is assumed to
prevent double counting. In this notation the initial state is written
as $\ket{1_0}=\ket{1/2,\{\},\{\},\{\}}$. In a similar fashion we can
also describe {the (exact)} 0-hole ground state.

An advantage of the introduced procedure, in which a variational space
grows like $N_{\rm st}\sim 6^M$ (see Tab.~\ref{var-tab1}), is that
\begin{table}[t]\begin{center}
\makebox{\vbox{\hrule depth 0.8pt
\hbox{\vspace{-1pt}}
\hbox{\begin{tabular}{c||rc||rccc}
\raisebox{-3mm}{$M$}
& \multicolumn{2}{c||}{\rule[3mm]{0mm}{2mm}0-hole system}
& \multicolumn{4}{c}{1-hole system}\\
& \rule[-2.5mm]{0mm}{2mm} $N_{\rm st}$ & $\Delta^{(0)}$
& $N_{\rm st}$ & $E_0^{(1)}$ & $\Delta^{(1)}$
& $\Lambda^{(1)}$
  \\ \hline\hline
  \rule[4mm]{0mm}{2mm}0
  & 3 &  & 2 &  &  &  \\[0.2cm]
  1 & 11 & 4.206 & 13 & 3.852 & 2.486 & 5.679 \\[0.2cm]
  2 & 59 & 3.507 & 78 & 3.525 & 1.976 & 3.423 \\[0.2cm]
  3 & 377 & 3.225  & 507 & 3.416 & 1.784 & 2.525 \\[0.2cm]
  4 & 2469 & 3.100 & 3354 & 3.379 & 1.690 & 2.100 \\[0.2cm]
  5 & 16186 & 3.042 & 22141 & 3.368 & 1.634 & 1.903 \\[0.2cm]
  6 & 105663 & 3.015 & 145402 & 3.364 & 1.604 & 1.820 \\[0.2cm]
  7 & 686111 & 2.998 & 948025 & 3.363 & 1.584 & 1.777 \\[0.2cm]
  8 & 4429695 & 2.983 & 6144089 & 3.363 & 1.576 & 1.762 \\[0.2cm]
  \vdots &  & \vdots &  & \vdots & \vdots & \vdots \\[0.2cm]
  $\infty$ &  & 2.979 & & 3.362 & 1.573 & 1.745 \\[0.1cm]
\end{tabular}}
\hrule depth .8pt}}
\caption{Variational calculation for $J=76.8$~K, $J'/J=0.62$. $M$
($N_{\rm st}$) denotes the number of variational steps (states) and
$\Delta^{(0)}=E^{(0)}_1(S=1)-E^{(0)}_0(S=0)$ represents the spin gap
of the undoped system. $E^{(1)}_0$ is the ground state energy of the
1-hole system measured relative to the ground state energy
$E^{(0)}_0(S=0)$ of the 0-hole system.
$\Delta^{(1)}=E^{(1)}_1(S=1/2)-E^{(1)}_0(S=1/2)$ and
$\Lambda^{(1)}=E^{(1)}_2(S=1/2)-E^{(1)}_0(S=1/2)$ where
$E^{(1)}_1(S=1/2)$ and $E^{(1)}_2(S=1/2)$ represent the first and
second excited states of the 1-hole system, respectively. We also show
extrapolated values for $M\to\infty$. All energies are listed in meV.}
\label{var-tab1}
\end{center}\end{table}
triplet dimers of newly generated variational states develop
gradually around a hole, which is very promising in a way that
only a few such states would suffice to correctly describe the
localized nature of the spin polaron found in
Refs.~\cite{p4-vojta,p4-leung} As application of the off-diagonal
part ($J'$) of Hamiltonian $H$ on inter-singlet-dimer bonds is
zero, the relevant region from where new triplets may be generated
is limited to a neighborhood of the hole. Due to specific topology
of the SS lattice, a gradual expansion of triplet region with
increasing $M$ is proceeded in a zig-zag path. A short calculation
shows that already $M=3$ is enough for the outermost triplet to
exceed the square region of 32 sites used in Ref.~\cite{p4-leung}.

Along with the construction of the variational space we calculated
nonzero matrix elements $H_{ij}^M=\innx{i_M}{H}{j_{M-1}}$ separately
for each term in Hamiltonian~(\ref{eq_shs0}). {Matrix} $H_{ij}^M$
was diagonalized at each generation $M$ to control the convergence.
\begin{figure}[t!]\begin{center}
    \includegraphics[width=0.38\textwidth]{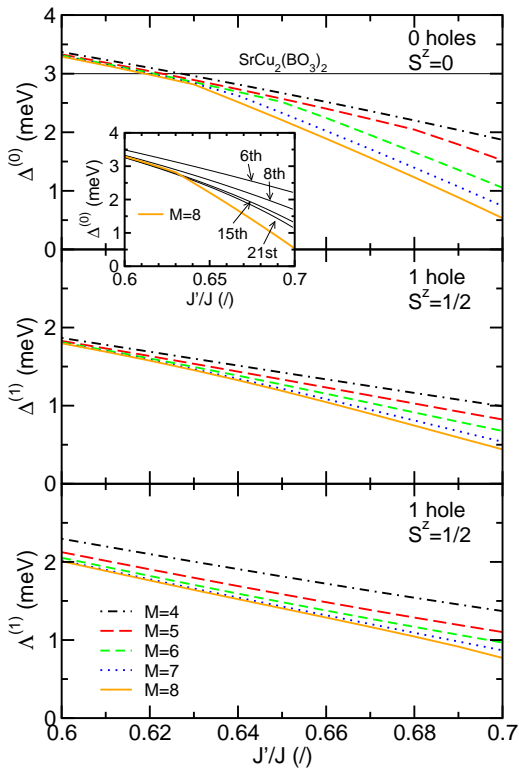}
    \caption{(color online) Energies relative to the ground-state
    energy vs $J'/J$ calculated at $J=76.8$~K for 0- and 1-hole
    system, defined on infinite SS lattice, using $M$ steps of the
    variational method. Black solid line represents experimental value
    obtained on SrCu$_2$(BO$_{3}$)$_{2}$. In the inset we show the
    comparison with the 6th, 8th, 15th and 21st order dimer-expansion
    calculation in $J'/J$ for 0 holes. \cite{w-zheng,s-knetter1}}
\label{fig_var_1}
\end{center}\end{figure}
In Fig.~\ref{fig_var_1} we show the convergence of a few lowest
eigenenergies for the 1-hole and, for comparison, also for the
0-hole case. Concentrating first on the single impurity case, we
plot the energy difference between the ground state and the first
two excitations, $\Delta^{(1)}$ and $\Lambda^{(1)}$, as a function
of $J'/J$ at fixed $J=76.8$~K.
%
%
In this particular calculation two initial state vectors were used
to improve the sampling of the variational space: $\ket{1_0}$ and
$\ket{2_0}=\ket{-1/2,\{\},\{(0,1)\},\{\}}$ with one $S^z=1$
triplet located above the uncoupled spin as also depicted in
Fig.\ref{fig_doped}(b). In addition, we choose to work in the
$S^z=1/2$ sector that contains the ground state of the 1-hole
system in the absence of the magnetic field. No other symmetries
were explicitly employed in the construction of the variational
space. The SS model doped with a static hole possesses  the mirror
symmetry over the plane containing a doped dimer. We have checked
that eigenstates and corresponding correlation functions transform
according to this symmetry.

Moving back to the convergence of results for the  1-hole case, as
presented in Fig.~\ref{fig_var_1}, we notice, that values of
$\Delta^{(1)}$ and $\Lambda^{(1)}$ have nearly converged for
$J'/J\lesssim 0.62$ for $M=8$, i.e., their values at $M=8$ are not
far from their extrapolated values for $M\to \infty$. As expected,
the variational approach becomes worse when $J'/J$ approaches the
transition to the AFM state of the 0-hole case, $J'/J\sim 0.7$.
This can be better seen in a 0-hole case where the spin gap
$\Delta^{(0)}$ should close as $J'/J\to 0.7$. In our calculations
{$J'/J\sim 0.64$ represents a critical ratio}
beyond which the convergence of the spin gap $\Delta^{(0)}$ is
lost. On the other hand, at $J'/J=0.62$ a single-triplet energy is
quite well converged for $M=8$, which in turn leads to
thermodynamically converged value of the spin gap $\Delta^{(0)}$ that
also agrees with the experimental value $\Delta^{(0)}\sim 3$~meV. This
suggests that the same set of $J$ and $J'$ as in calculations on 20
sites \cite{p2-jorge03,p3-elshawish,p4-elshawish} may be used here in
order to compare with realistic SrCu$_{2}$(BO$_{3}$)$_{2}$ system.  We
attribute this fast convergence to the thermodynamic limit to the fact
that single-triplet excitations are well localized, which renders a
limited variational space sufficient for nearly exact description of
low-energy excitations.

A 0-hole case may also serve as a test for the method, for it can be
directly compared to alternative approximate approaches. A comparison
with dimer-expansion technique of Zheng {\it et al.} \cite{w-zheng}
and Knetter {\it et al.} \cite{s-knetter1} is shown in the inset of
Fig.~\ref{fig_var_1}. Our method for $M=8$ gives similar energy-gap
dependence as the 21st order calculation for $J'/J\lesssim 0.64$,
however, at larger values of $J'/J\gtrsim 0.64$ our method captures
better the closing of the spin gap. It should be noted at this point
that even better convergence with $M$ is expected when a translational
symmetry of the lattice is employed for the 0-hole case.
%
\section{Results}
%
\begin{figure}[t!]\begin{center}
    \begin{minipage}[b]{0.45\textwidth}\begin{center}
        \includegraphics[width=\textwidth]{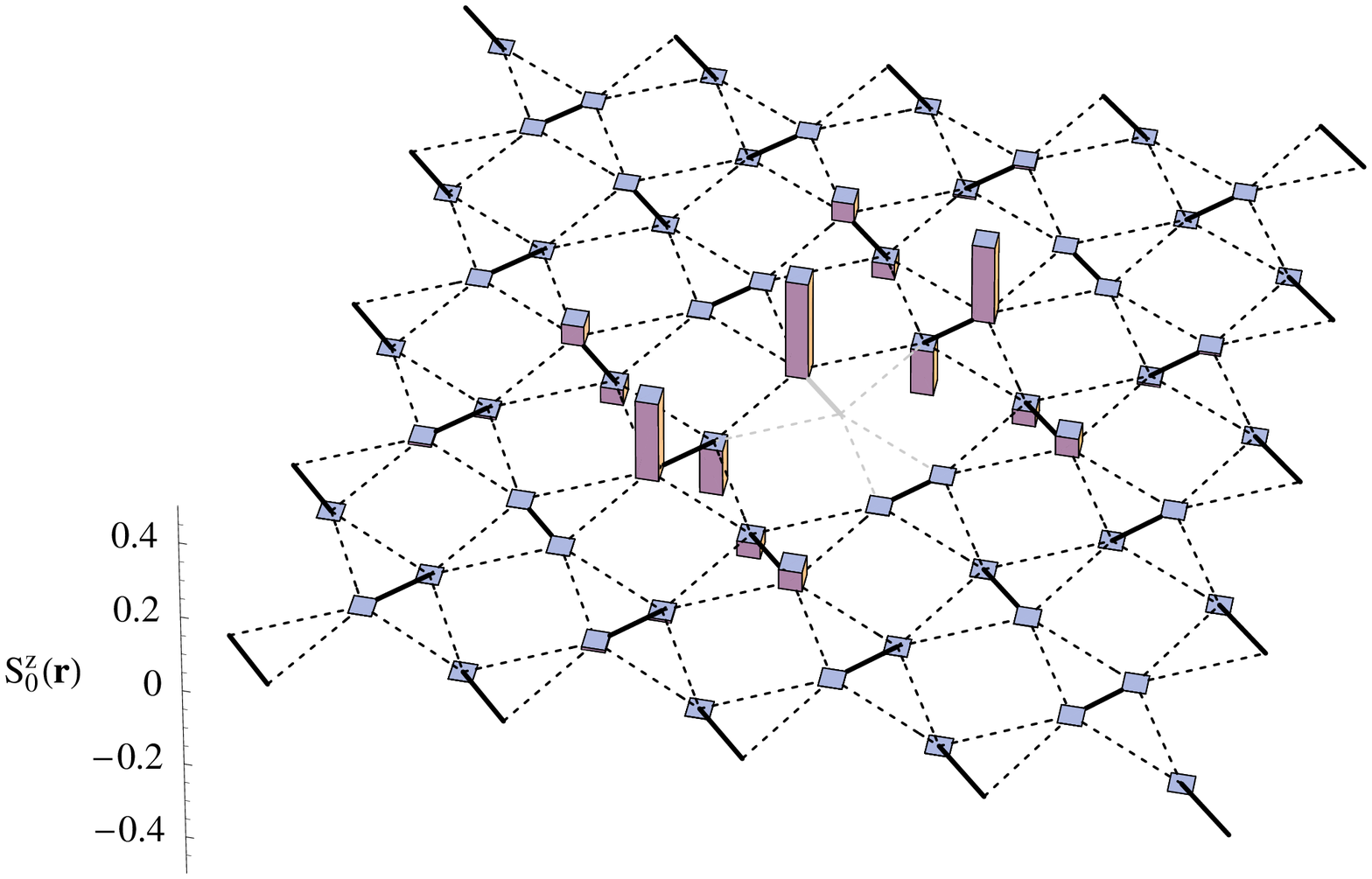}
        \vskip-0.3cm (a)
    \end{center}\end{minipage}
    \vskip0.3cm
    \begin{minipage}[b]{0.45\textwidth}\begin{center}
        \includegraphics[width=\textwidth]{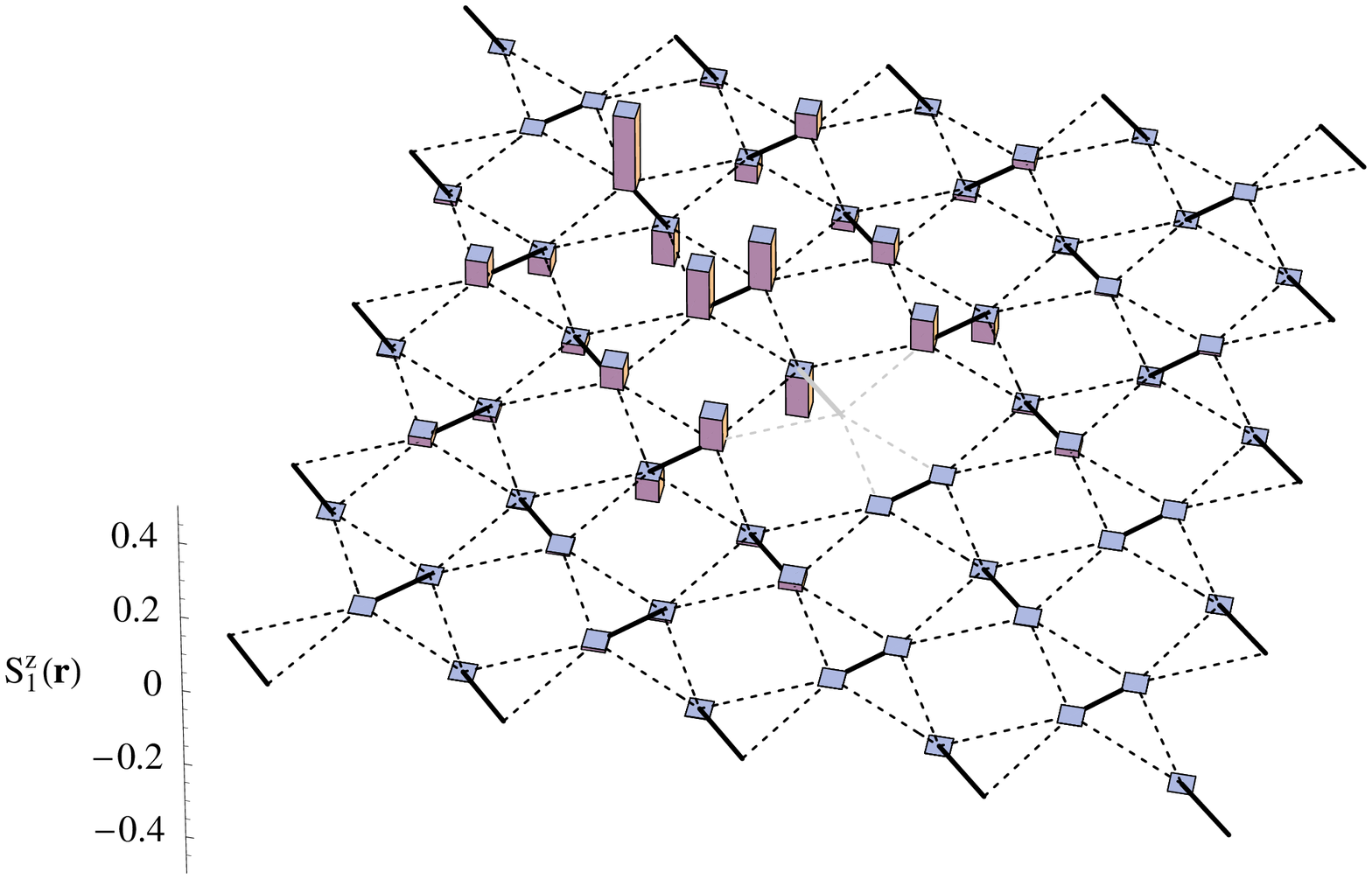}
        \vskip-0.3cm (b)
    \end{center}\end{minipage}
    \vskip0.3cm
    \begin{minipage}[b]{0.45\textwidth}\begin{center}
        \includegraphics[width=\textwidth]{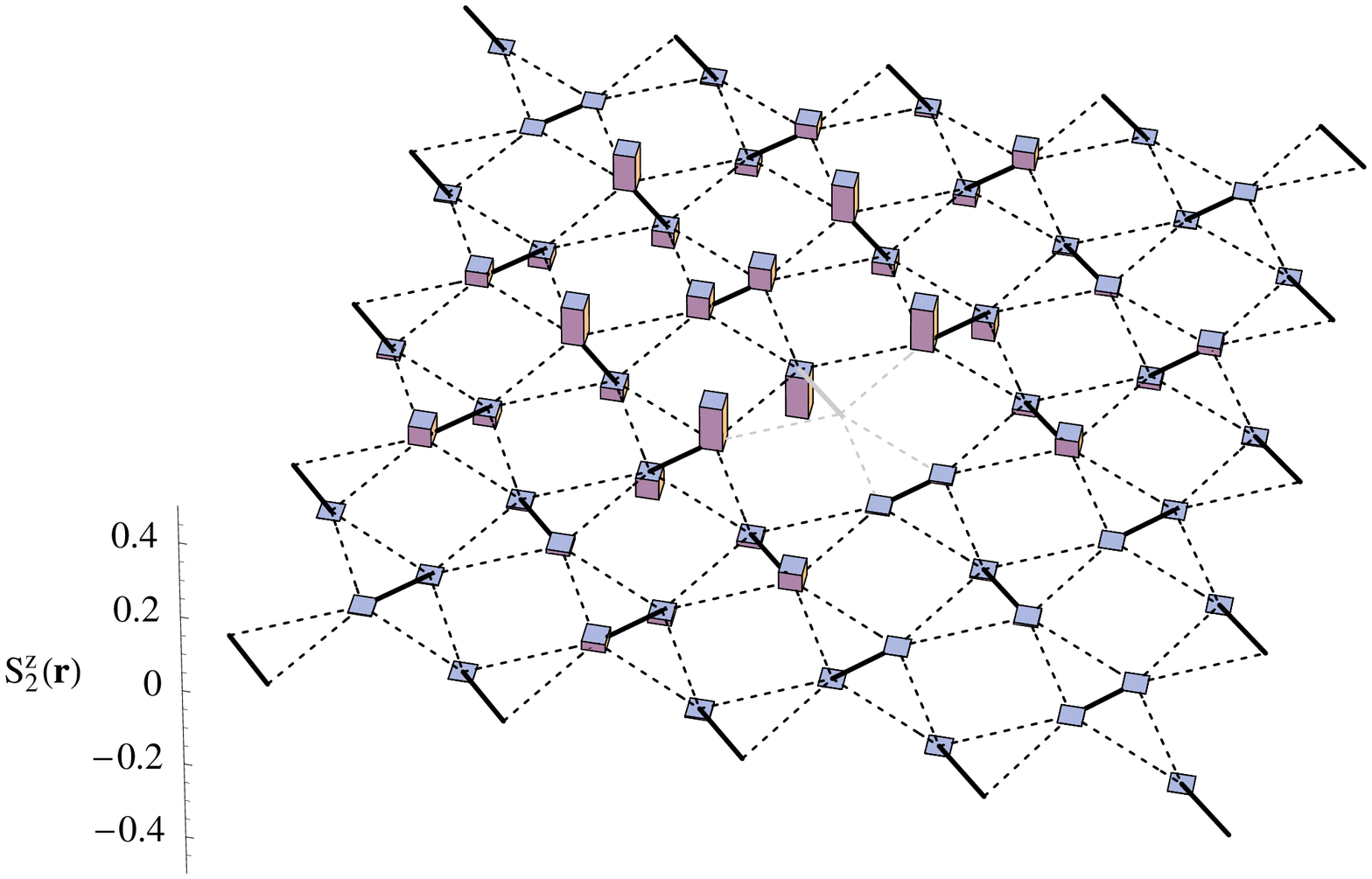}
        \vskip-0.3cm (c)
    \end{center}\end{minipage}
    \caption{(color online) Spin distribution function $S_n^z({\bf
    r})$ calculated in (a) ground state, (b) first excited state, and
    (c) second excited state of the 1-hole SS model (\ref{eq_shs0}).
    Calculation was performed for $J=76.8$~K, $J'/J= 0.62$ and $M=8$.}
\label{fig_polaron}
\end{center}\end{figure}
We have calculated the shape of a single spin polaron, forming in
the vicinity of a static hole, as measured by the spin
distribution function calculated in the 1-hole (by hole we denote
the magnetic impurity) eigenstate $\ket{n^{(1)}}$ of $H$
(\ref{eq_shs0}):
\begin{equation}
   S_n^z({\bf r})=\innx{n^{(1)}}{S_{{\bf r}}^z}{n^{(1)}},
   \hbox{\hspace{10mm}}n=0,1,2.
\end{equation}
Here ${\bf r}$ denotes the position of the spin with respect to the
hole. We first note that in the case of $J'/J \lesssim 0.7$ and zero
doping $S_0^z({\bf r})=0$ for all ${\bf r}$ due to a singlet nature of
the ground state $\ket{0^{(0)}}$.

In Fig.~\ref{fig_polaron} we present the shape of the spin polaron,
$S_0^z({\bf r})$, and of its two lowest excitations, $S_1^z({\bf r})$
and $S_2^z({\bf r})$, calculated for $M=8$.
The zero value of $S_n^z({\bf r})$ indicates that ${\bf r}$ points
either towards the edges of a singlet or a $S^z=0$ triplet dimer.
Note also that the following sum-rule holds: $\sum_{\bf r}
S_n^z({\bf r})=1/2$.
From Fig.~\ref{fig_polaron} it is apparent that the ground state
of the spin polaron is rather well localized around the impurity.
Only first two neighbors lying in the direction perpendicular to
the dimer containing the impurity are strongly affected by the
presence of the impurity. The other two neighbors lying along the
direction of the dimer with the impurity are nearly unaffected
giving $S_0^z({\bf r})\approx 0$. The shape of the spin polaron is
therefore extremely anisotropic. This is in agreement with the
exact $T=0$ calculations on 32 sites. \cite{p4-leung} According to
Ref.~\cite{p4-leung}, a mobile hole introduced to the dimerized
ground state forces the resulting free spin to minimize the energy
by forming a five-spin chain with two of its nearest dimers. This
excitation is localized and has a small dispersion compared to the
$t$-$J$ model on a square lattice. These results are in
qualitative agreement with our small-spin-polaron picture around
the localized hole.
\begin{figure}[t!]\begin{center}
    \includegraphics[width=0.38\textwidth]{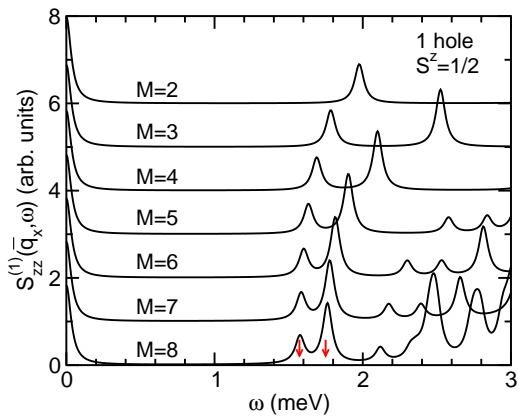}
    \caption{(color online) $q_x$-averaged 1-hole dynamical spin
    structure factor $S_{zz}^{(1)}(\bar{q}_x,\omega)$ as a function of
    $M$ calculated for $J=76.8$~K and $J'/J=0.62$. Lines are shifted
    for clarity. The arrows denote extrapolated positions of the
    lowest two peaks.}
\label{fig_var_2}
\end{center}\end{figure}

In Fig.~\ref{fig_polaron} we also show a spatial distribution of
the lowest two excited states of the spin polaron, $S_1^z({\bf
r})$ and $S_2^z({\bf r})$, calculated as well in $S^z=1/2$ sector
and for $M=8$. In contrast to the $S_0^z({\bf r})$ case,
$S_1^z({\bf r})$ shows spin disturbance predominantly along the
direction of doped dimer. Intriguingly, this excitation is not
centered around the impurity; instead it is shifted away from it
towards the neighboring orthogonal dimer. We note, however, that
this excited state of the polaron is not confined by the extent of
the variational space which allows displacements at much larger
distances. This leads to the conclusion that the lowest polaron
excitation is centered on the neighboring orthogonal dimer, yet it
remains localized in the vicinity of the hole. A {slightly}
different picture is obtained for $S_2^z({\bf r})$ where the
uncompensated spin is distributed more evenly on NN and NNN around
the hole. As naively expected, higher energy excitations of the
spin polaron spread further away from the impurity site.  For this
reason more variational states are needed to obtain full
convergence in this case.
\begin{figure}[t!]\begin{center}
    \includegraphics[width=0.38\textwidth]{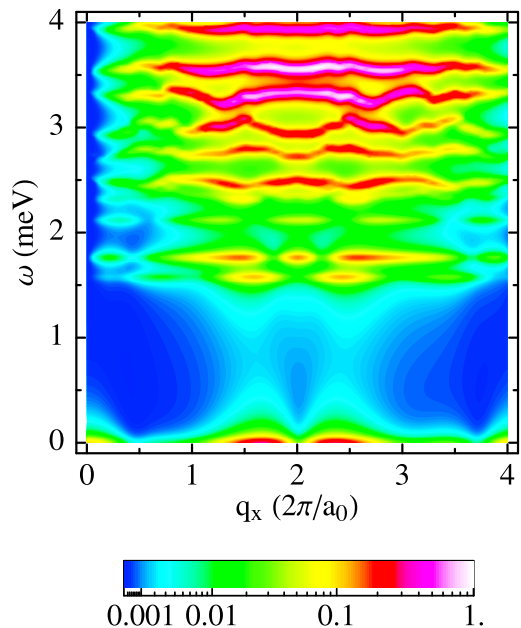}
    \caption{(color online) Intensity plot of the 1-hole dynamical
    spin structure factor $S_{zz}^{(1)}({\bf q},\omega)$ calculated
    for $J=76.8$~K, $J'/J=0.62$ and $M=8$. A map was obtained by
    interpolation between 40 equidistant values of $q_x\in [0,4]$.
    Note also the logarithmic intensity scale.}
\label{fig_surf}
\end{center}\end{figure}

In the last part we investigate the effect of spin polaron states,
forming around static nonmagnetic impurities, on  {scattering
experiments}. In Fig.~\ref{fig_var_2} we present the dynamical
spin structure factor calculated within our variational approach
at finite wavevector ${\bf q}$,
\begin{eqnarray}
   S_{zz}^{(1)}({\bf q},\omega)
   &=&-{\rm Im}\innxx{0^{(1)}}{S_{{\bf q}}^z\,\frac{1}{\omega^+ +E_0^{(1)}-H}\,
   S_{-{\bf q}}^z} {0^{(1)}}; \nonumber \\
   S_{{\bf q}}^z&=&\sum_{{\bf R},{\bf r}}
   {\rm e}^{{\rm i}{\bf q}({\bf R}+{\bf r})}S_{{\bf R},{\bf r}}^z.\label{som2}
\end{eqnarray}
Here ${\bf R}$ runs over all unit cells reached within $M$ steps and
${\bf r}$ spans four vectors forming the basis of the unit cell that
contains two orthogonal dimers. For the details describing interatomic
distances in SrCu$_{2}$(BO$_{3}$)$_{2}$ system we refer the reader to
Ref.~\cite{p3-knetter}. In addition, a finite $\epsilon=0.25$~meV
($\omega^+=\omega+{\rm i}\epsilon$) is used to smooth the spectra.

%
%
In Fig.~\ref{fig_var_2} we show spectra of
$S_{zz}^{(1)}(\bar{q}_x,\omega)$ averaged over the interval of $q_x\in
[0,4]$ (in units of $2\pi/a_0$) for various $M$. It is evident that
spectra at $\omega\lesssim 2$~meV has fully converged, while
convergence at higher frequencies is slower.  Nevertheless, from
Fig.~\ref{fig_var_2} it is apparent that finite doping leads to a
finite response of the dynamical spin structure factor at frequencies
below the gap value, $\omega\approx 3$~meV, of the undoped
system. Similar effect is as well seen in the surface plot presented
in Fig.~\ref{fig_surf} where we explicitly provide the $q_x$
dependence that was computed using 40 discrete values evenly spaced on
the interval $q_x\in [0,4]$.
Despite slower convergence at higher frequencies, as observed in
Fig.~\ref{fig_var_2}, two predictions can be made based on
presented calculations. We first predict that in-gap peaks should
appear at frequencies far below the gap value in the {scattering}
experiment on a doped SrCu$_{2}$(BO$_{3}$)$_{2}$ compound, where
some Cu atoms would be replaced with static nonmagnetic impurities
(holes). These in-gap peaks would carry information about the
excited spin-polaron states. Second, doped impurities would
contribute to broadening of the one-magnon peak at $\omega\approx
3$~meV through inelastic scattering of magnons on excited states
of spin polarons surrounding the impurities. This effect can be
clearly seen from Figs.~\ref{fig_var_2} and \ref{fig_surf} where
extra peaks appear in the vicinity of $\omega\approx 3$~meV in
comparison with calculations on the undoped system.
\cite{p4-elshawish,s-knetter1}

%
\section{Conclusion}
%
In summary, we presented static and dynamic properties of a
{single} spin polaron in a doped SS lattice. We have developed a
variational approach, defined on the infinite lattice, that gives
numerically highly accurate results in the thermodynamic limit of
static properties of spin polaron and its lowest excitations. The
first and second excited state of spin polaron represent $S=1/2$
excitations, their energies lie well within the spin gap of the
undoped system. The shape of the spin polaron is highly
anisotropic while its extent is well localized in agreement with
exact diagonalizations on finite systems. \cite{p4-leung} Spin
polaron extends predominantly to the first neighboring dimers
lying in perpendicular direction with respect to the dimer
containing the impurity. The shape of the first excited state of
the polaron extends further in space and is off-centered with
respect to the position of the impurity. Spin structure factor of
the doped {SS model}
%
%
is consistent with the appearance of in-gap peaks and the broadening
of the one-magnon peak due to inelastic scattering of the magnon on
static spin polarons.


%
\acknowledgments
We acknowledge the inspiring discussions with B.~D. Gaulin that took
place during the preparation of this manuscript. We also acknowledge
the financial support of Slovene Research Agency under contract
P1-0044.
%

%
\end{document}